\documentclass[12pt]{spieman}  
\usepackage{amsmath}
\usepackage{amssymb}
\usepackage{amsfonts}
\usepackage{graphicx}
\usepackage{setspace}
\usepackage{tocloft}
\usepackage{xcolor}
\usepackage{bbm}
\usepackage{subfig}
\usepackage[countmax]{subfloat}
\usepackage{framed}
\usepackage{color}
\usepackage{hyperref}
\usepackage{epstopdf}
\usepackage{dsfont}
\usepackage{wrapfig}
\usepackage{relsize}
\usepackage{bm}
\usepackage{enumitem}
\usepackage{soul}
\usepackage{physics}

\title{Fast remote spectral discrimination through ghost spectrometry}

\author[a]{Andrea Chiuri}

\affil[a]{ENEA - Centro Ricerche Frascati, via E. Fermi 45, 00044 Frascati, Italy}

\author[b,c]{Marco Barbieri}

\affil[b]{Dipartimento di Scienze, Universit\`a degli Studi Roma Tre, Via della Vasca Navale 84, 00146 Rome, Italy}
\affil[c]{Istituto Nazionale di Ottica, CNR, Largo Enrico Fermi 6, 50125 Florence, Italy}

\author[b]{Iole Venditti}

\author[a]{Federico Angelini}

\author[b]{Chiara Battocchio}

\author[d,e]{Matteo G A Paris}

\affil[d]{Department of Physics "A. Pontremoli", Università degli Studi di Milano, I-20133 Milano, Italy}
\affil[e]{Istituto Nazionale di Fisica Nucleare, Sezione di Milano, 20133 Milano, Italy}

\author[b,*]{Ilaria Gianani}

\cftpagenumbersoff{figure}
\cftpagenumbersoff{table} 
\begin{document} 
\maketitle

\begin{abstract}
Assessing the presence of chemical, biological, radiological and nuclear threats is a crucial task which is usually dealt with by analyzing the presence of spectral features in a measured absorption profile. The use of quantum light allows to perform these measurements remotely without compromising the measurement accuracy through ghost spectrometry. However, in order to have sufficient signal-to-noise ratio, it is typically required to wait long acquisition times, hence subtracting to the benefits provided by remote sensing. In many instances, though, reconstructing the full spectral lineshape of an object is not needed and the interest lies in discriminating whether a spectrally absorbing object may be present or not. Here we show that this task can be performed fast and accurately through ghost spectrometry by comparing the low resources measurement with a reference. We discuss the experimental results obtained with different samples and complement them with simulations to explore the most common scenarios. 
\end{abstract}

{\noindent \footnotesize\textbf{*}Ilaria Gianani:  \linkable{ilaria.gianani@uniroma3.it} }

\begin{spacing}{2}   

\section{Introduction}
 
Spectroscopic techniques are a fundamental tool for the characterization of materials~\cite{pavia2014introduction}. For centuries these have been successfully employed in a variety of fields and have been diversified to account for the most diverse scenarios and needs~\cite{koral2021thz,hammes2005spectroscopy,adriaens2021spectroscopy,appenzeller2012introduction,schaefer1998spectroscopic,wagatsuma2021spectroscopy}. Including the new capabilities enabled by quantum light has widened the already vast range of possible applications, in particular for what concerns harnessing quantum frequency correlations~\cite{Mukamel20}. In recent years, two main routes have been pursued. The first exploits spectral correlations between two photons in non-degenerate configurations~\cite{Kalachev_2008}, so that hardly accessible spectral regions can be explored by looking at their correlated counterpart in the visible range~\cite{Chan09,Kalashnikov16,Paterova20}. The second employs the correlation to perform remote sensing measurements~\cite{PhysRevX.4.011049}, akin to ghost imaging protocols~\cite{Aspden15,Pepe16,Ryczkowski16,lemos14nat,Moreau18}. This latter route can be particularly advantageous when the objects at hand are not easily accessible or represent so-called chemical, biological, radiological, and nuclear (CBRN) threats \cite{chierici20}. In these instances it is vital to extract information at a distance, both to ensure the safety of the users and for ease of measurement operations. Although this is also possible using classical spectral correlations, using quantum ones can show a better performance~\cite{sullivan10pra,bennink04prl,padgett17ptrsa}, especially when the number of modes to be considered is large~\cite{PhysRevA.105.013506}. 

Dangerous components are typically recognised by features in the absorbance spectrum, however, in order to fully retrieve the lineshape of such spectral objects a good signal to noise ratio (SNR) is desirable and, given the typical efficiencies, this usually results in long accumulation times times. This poses a strong limitation and dramatically hinders the benefits arising from the use of quantum resources. 

If we are interested in swiftly assessing the presence of a threat, retrieving its full lineshape may not be necessary. One may, in fact, recast the problem as a discrimination one, and wonder whether it is possible to infer the presence of the threat comparing a fast low-signal spectral measurement performed on the supposed threat, with a reference measurement. A common technique for discrimination makes use of the correlation coefficient between vectors representing the spectra, or, alternatively, of their distance~\cite{ghorbani2019mahalanobis}. These are versatile tests, since no requirements on the distribution are needed; on the other hand, these are prone to artefacts at low SNRs leading to the wrong attribution. 

A decision procedure (an {\em inference strategy}) prescribes which hypothesis has to be chosen given a set of data. Then, one assigns a {\em cost} to the choice of the null hypothesis ({\it e.g.} no threat) when the alternative hypothesis is true and look for a strategy minimizing the average cost.  In a Bayesian approach, one assigns equal cost to any wrong inference and zero cost to correct one, such that the average cost equals the overall probability of error. This approach has been applied to spectroscopy with success~\cite{Silva92}, but a fully Bayesian approach for large set of data may be challenging. One rather employs the concept of likelihood ratios to evaluate the posterior probabilities.

For binary discrimination problems where the alternative hypothesis has a low a priori probability to occur (i.e. when the threat is not likely to be present) one may also employ the so-called Neyman-Pearson strategy instead of a Bayesian one \cite{Paris19}. The optimal NP strategy maximizes the probability of revealing the threat when it is there instead of minimizing the probability of error \cite{scott}. Following these ideas,  we tackle the problem of risk detection by performing a Kolmogorov-Smirnov test (KST) \cite{razali2011power} between a high SNR reference and a low SNR measurement with a ghost spectrometer.The Kolmogorov-Smirnov test was specifically developed to test the hypothesis that two samples come from the same distribution and still represents one of the most powerful non-parametric tools of hypothesis testing. 

We demonstrate an experiment on two different targets and complement our results by simulating  different operational regimes. Our results show that even with slightly absorbing spectral objects this techniques allows to ascertain the presence of an object with a limited number of resources and requiring limited processing on the data, thus enabling time-efficient discrimination.

\begin{figure}
    \centering
    \includegraphics[width=\textwidth]{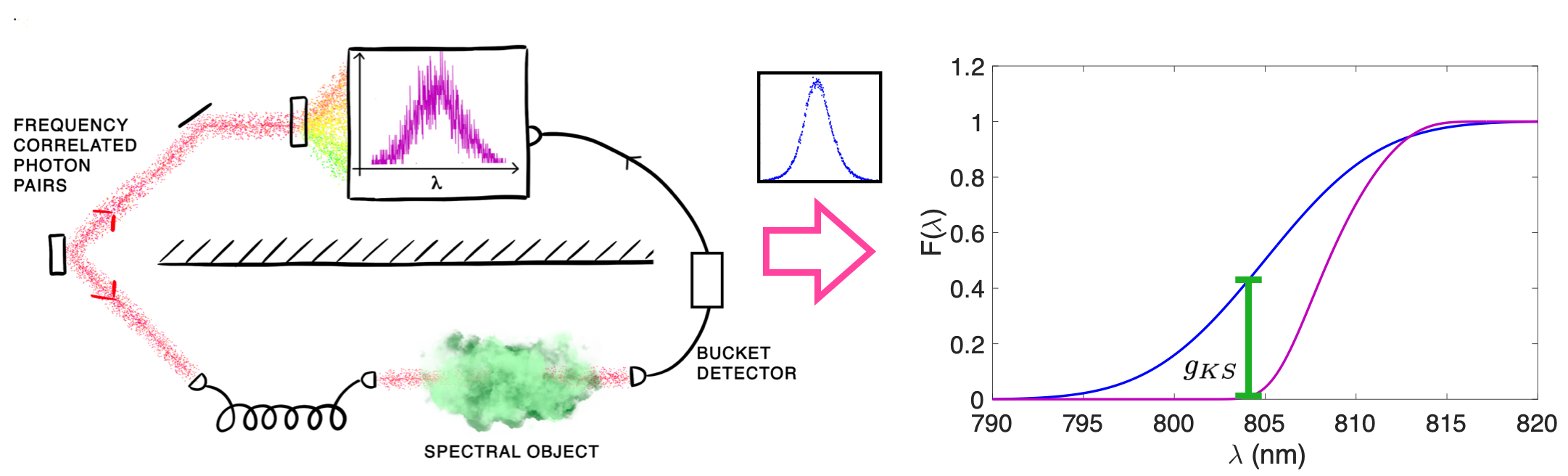}
    \caption{Conceptual scheme. A ghost-spectrometry scheme is implemented using frequency-correlated photon pairs: one photon is directed towards the supposed location of the spectral object, and detected with a bucket detector, while the other photon is analyzed in frequency. By comparing the recorded coincidences at low SNR with a reference measurement using a Kolmogorov-Smirnov test it is possible to asses if the two profiles are samples drawn from the same distribution or not.  }
 
    \label{fig:scheme}
\end{figure}

\section{Method}

Our objective is that of remotely discriminating between the presence or absence of an object that has an absorption profile in a given spectral range. The object could be for instance a CBRN threat that needs to be identified swiftly and whose location cannot be easily accessed.

To perform the measurement remotely we consider the quantum ghost spectrometer scheme shown in Fig. \ref{fig:scheme}: frequency-correlated photon pairs are generated in the region of interest. One photon is sent at the object location and is detected with a bucket detector, while the other photon is detected locally with a spectrally-resolved measurement. In this way, the demanding spectral measurement can be performed far from the object location, thus allowing for ease of operations.  
We use the same scheme to perform two measurements: first we perform a measurement with high SNR without any spectral object. We thus record the spectral distribution $S_r$ which will serve as a reference. We then perform a measurement of the signal at the location where the spectral object should be, and record the transmitted spectrum with a low SNR $S_s$. If there were no spectral object, $S_s$ and $S_r$ would have to be two samples drawn from the same distribution. This constitutes our null hypothesis. We can then perform a Kolmogorov-Smirnov test to accept or reject this hypothesis. In order to perform the KST, one proceeds as follows: starting from the two measured profiles for the reference and the signal one builds the two cumulative distributions $F_r$ and $F_s$ and evaluates the quantity: 
\begin{equation}
    g_{KS}= \max_{\lambda} \vert F_s(\lambda)-F_r(\lambda)\vert,
\end{equation}
which is a statistical variable of known distribution. Depending on its value one can either reject of accept the null hypothesis. 
In Fig \ref{fig:scheme} we show an example of the cumulative distributions for reference (blue) and signal (purple). $g_{KS}$ identifies the maximum separation between the two.  

\begin{figure}
    \centering
    \includegraphics[width=\textwidth]{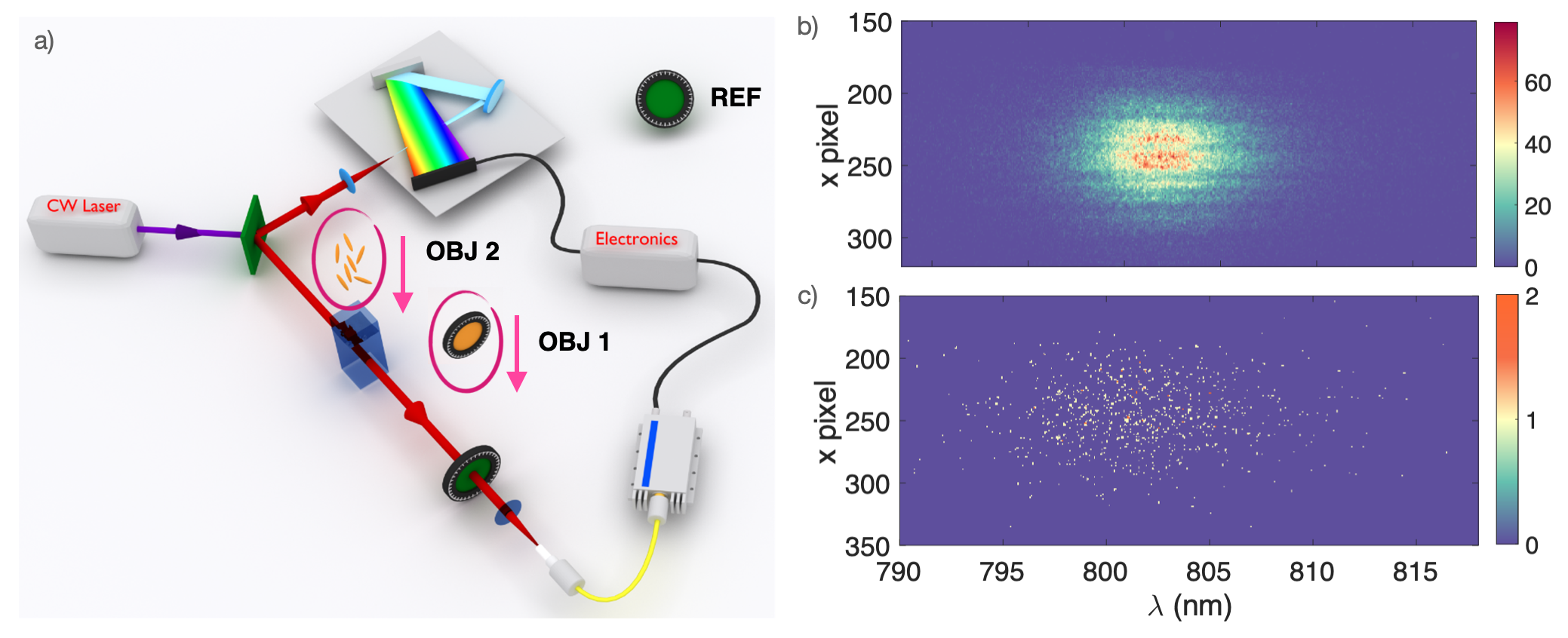}
    \caption{ a) Experimental Setup: a CW 403 nm laser is used to generate photon pairs through SPDC with a 3mm type I BBO crystal. One photon is sent through a spectrometer and detected with a iCCD camera, while the other photon is sent through a spectral object (either a supergaussian filter (orange) or a solution of AuNrs) and is detected using a bucket detector. A bandpass filter (green) acts as reference. b) Recorded coincidences for the reference. c) Example of recorded coincidences for AuNrs at an accumulation time t=10s.}
    \label{fig:setup}
\end{figure}

\section{Experiment}

We demonstrate this approach by performing measurements with the setup shown in Fig. \ref{fig:setup} a). We employ a CW laser at 403 nm to generate photon pairs through spontaneous parametric down conversion (SPDC) using a 3 mm thick Type I BBO crystal. One photon is detected locally with a spectrally-resolved measurement. This is performed with a spectrometer (Andor Kymera 328i) and a iccd (Andor iStar DH334T-18U-73) after 20 m of fiber to compensate for the camera delays. The other photon is directed towards the spectral object and is then detected with an avalanche photodiode (APD). A Gaussian bandpass filter centered at 810 nm selects the spectral region, thus acting as a reference. The camera is triggered by the signal coming from the APD and thus directly records the coincidences counts. Fine-tuning of the temporal overlap between the two arms is achieved by means of a FPGA. Panels (b) and (c) of Fig. \ref{fig:setup} show two examples of detected coincidences for the reference (b) and signal (c).  In order to retrieve the spectral profiles these are integrated over the spatial axis, selecting a region of interest. Note that the spectral axis refers to the wavelength of the photon correlated to that interacting with the spectral object and with the reference filter.

\begin{figure}
    \centering
   \includegraphics[width=0.8\textwidth]{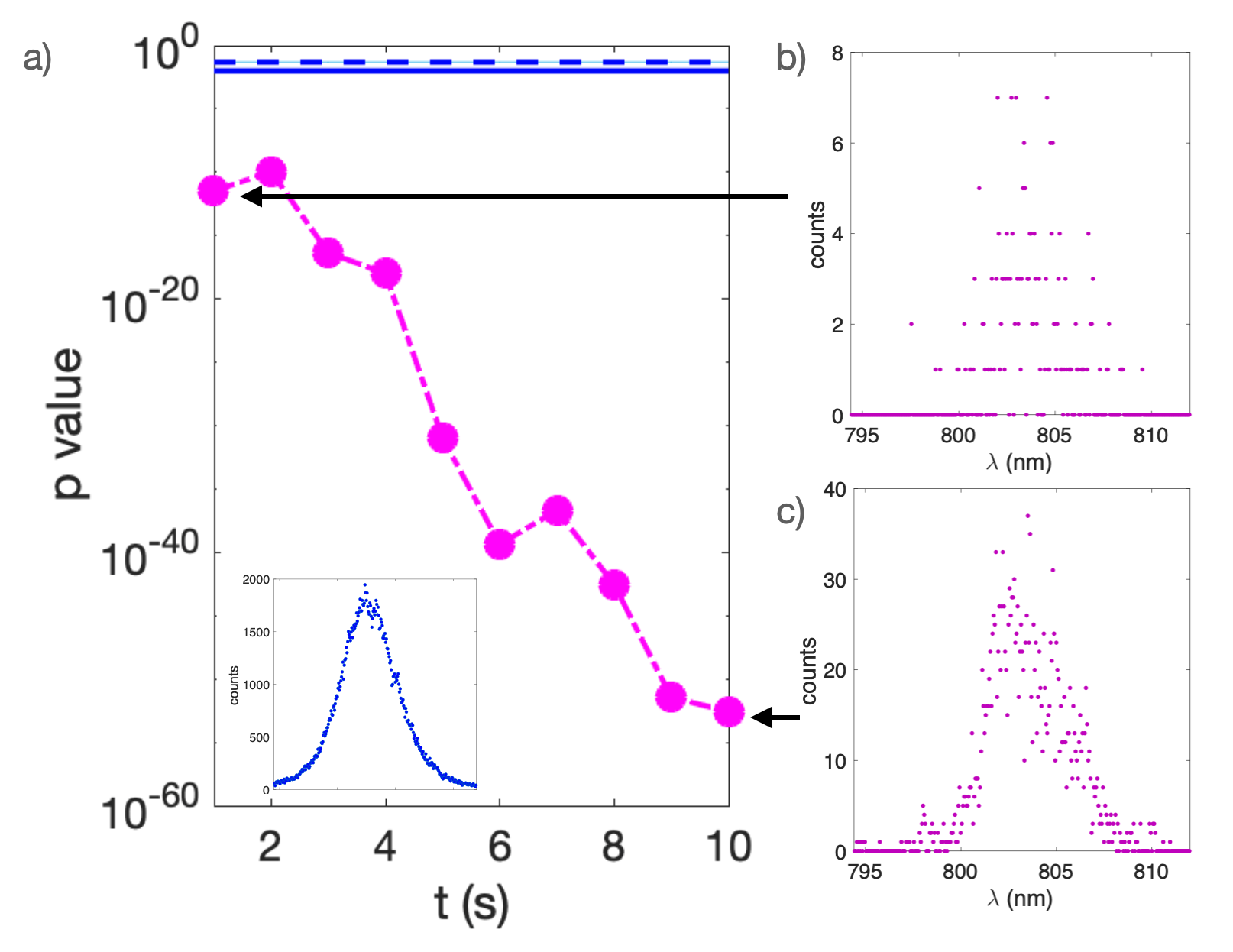}
    \caption{{\bf Results with supergaussian filter.} a) measured p values obtained from a KS test between the reference (blue inset) and the signal measured with the supergaussian filter inserted in the beam at different accumulation times t. The blue dashed and solid lines are the rejection confidence level at 0.05 and 0.01 respectively. b) and c) recorded coincidence counts at t=1 s and t= 10 s}
    \label{fig:exp1}
\end{figure}

We first test our technique by using as a spectral object a bandpass 4th-order super-Gaussian filter centered at 807 nm with a FWHM of 7.5 nm. We collect the reference spectrum performing a measurement without the object inserted, with a long accumulation time (t=600 s) so to achieve a good signal to noise ratio. The reference profile is shown in the inset of panel a) of Fig. \ref{fig:exp1}. This is the calibration of the system and, in case of threat detection, it can be performed in a safe environment.

Then we insert the filter and collect measurements with the accumulation time varying from 1 s to 10 s. Panels b) and c) of Fig. \ref{fig:exp1} show the profiles collected at t=1 s and t=10 s. We use the measured profiles to perform the KST, and report the obtained p values in Fig. \ref{fig:exp1} a) as a function of the accumulation time. Even for the signal at t=1 s, which corresponds to a total of 228 detected photons, we are able to reject the null hypothesis with a p value of $3\cdot 10^{-12}$. While this is remarkable given the limited counts required for a successful discrimination, the  profile of the object and the reference do differ significantly, having two different shapes and being centered at different wavelengths. This is not necessarily the case in a general scenario, where the spectral object may introduce more subtle discrepancies between the reference and the signal. 

\begin{figure}
    \centering
    \includegraphics[width=\textwidth]{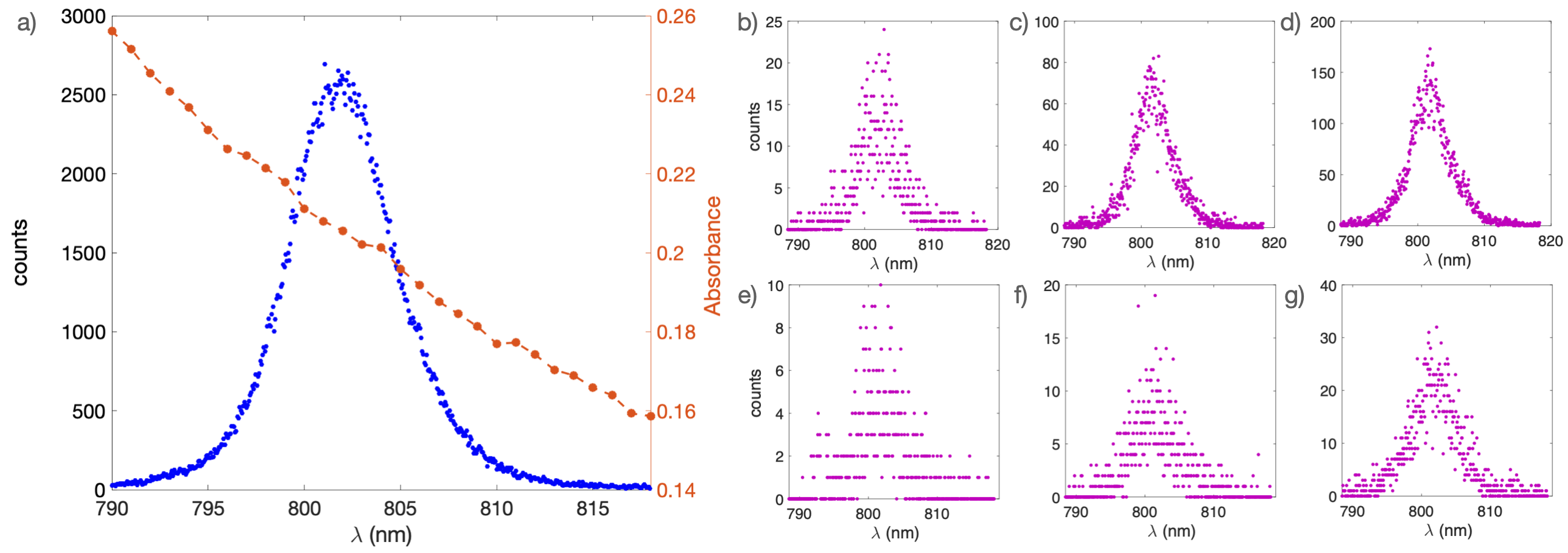}
    \caption{{\bf Nanorod profiles} a) Reference measurement (blue) and AuNRs absorbance (orange), (b-d) C1 coincidence mearuements at t=5,25,50, (e-g) C2 coincidence measurements at t= 1, 5,10. b) and e) correspond to a rejection rate of 0.3, c) and f) to a rejection rate of 0.7 for C1 and 0.8 for C2,  d) and g) to a rejection rate of 0.95.}
    \label{fig:exp2s}
\end{figure}

\begin{figure}
    \centering
    \includegraphics[width=\textwidth]{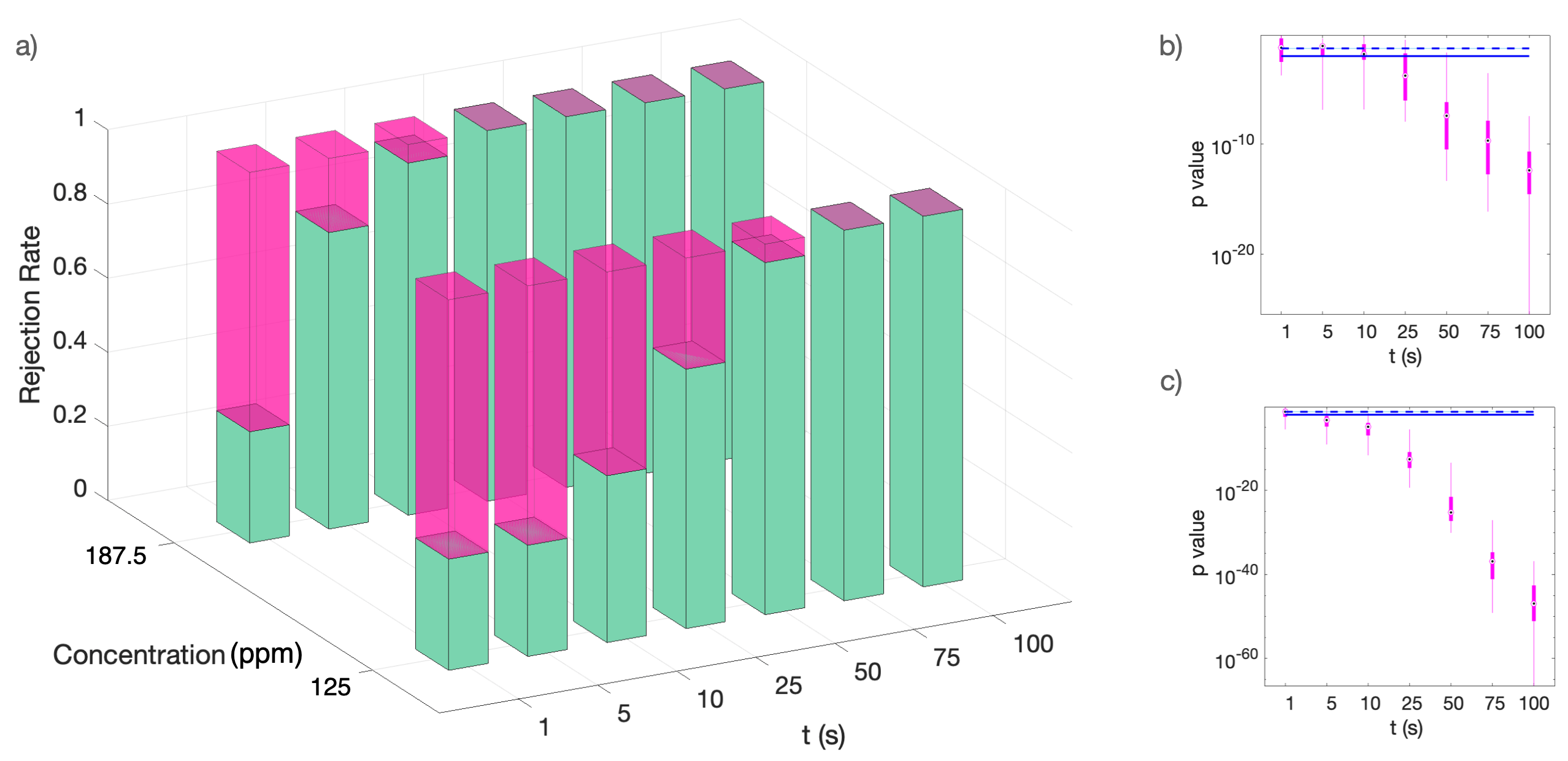}
    \caption{Results AuNRs: a) KS Rejection rate for concentration C1 and C2. Green bars indicate the successful rejection. Pink bars indicate acceptance of the null hypothesis. b) p values for C1, c) p values for C2.  Blue dashed line: 0.05 confidence level; blue solid line: 0.01 confidence level; the box center indicates the average value, while the box edges indicate the 25th and 75th percentile; the whiskers extend to all measured values.}
    \label{fig:exp2r}
\end{figure}

For this reason, we consider a second spectral object, i.e. a solution of gold nanorods (AuNRs) with a broad surface plasmon resonance band at 695 nm. The reference spectrum, being at the tail of the resonance band, will only experience a slight change in absorption with the wavelength (Fig. \ref{fig:exp2s}), so that the spectral distribution will differ from the reference considerably less than in the previous case. 
The AuNrs solution is placed in a quartz cuvette with a 1 cm path length. We perform the measurement for two different concentrations, 125 ppm (C1) and 188 ppm (C2), at different accumulation times ranging from t=1 s to t=100 s. In Fig. \ref{fig:exp2s} we report the profiles for the two concentrations measured at different times. For each accumulation time we record 20 measurements, and for each measurement we run the KST against a reference passing through distilled water only. The results are shown in Fig. \ref{fig:exp2r}. In panel a) we report the rejection rate (green) for the two concentrations at different accumulation times, normalized over the 20 measurements. 

The higher concentration results in a spectral distribution which will differ more from the reference compared to the lower concentration. This means that less resources are required for a successful discrimination. On the other hand, the lower concentration is more transparent, hence more resources will be collected per accumulation time. Even by taking this into account, while the rejection rate for C2 reaches  $100\%$ at t=25 s (corresponding to 3000 detected photons), for C1 the same is achieved at t=75 s (corresponding to 24000 detected photons). This is reflected in the measured p-values, which are shown in panel b) and c) of Fig. \ref{fig:exp2r} for C1 and C2 respectively. 

\begin{figure}[h!]
    \centering
    \includegraphics[width=\textwidth]{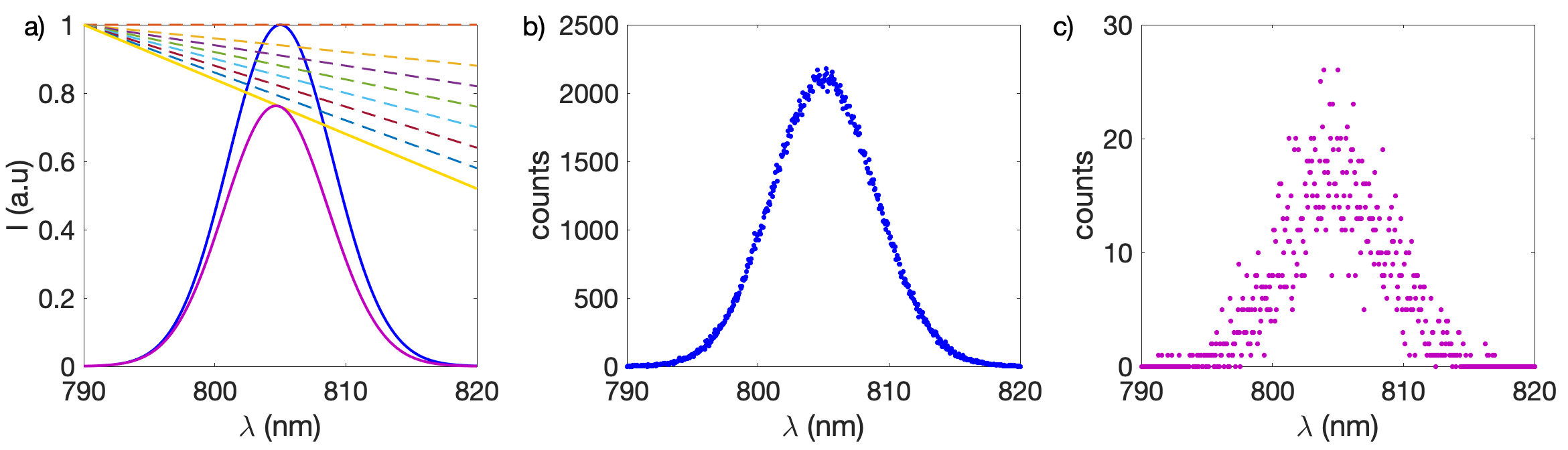}
    \caption{Simulation broad absorption on a structured reference. a) blu: reference, yellow and dashed lines: spectral object transmission, purple: signal obtained with the yellow transmission profile.  b) simulated reference c) simulated signal for $\alpha=0.016$ at the level of signal resulting in a $100\%$ rejection rate.   }
    \label{fig:sim1s}
\end{figure}
\section{Simulations}

In order to investigate the performance of our approach under typical regimes of operation, we complement the experimental results with numerical simulations. We explore two different scenarios: to provide an ideal benchmark to the example just discussed, we first look at the instance in which the absorption is much broader compared to the spectral region where the reference lies. Such broad spectra usually occur in UV-VIS spectroscopy~\cite{Perkampus12}. We then explore the regime in which the absorption is a narrow line compared to the reference region, as this is the most common occurrence when looking for narrow peaks in the fingerprint region of a IR~\cite{Griffiths07} or in a Raman spectrum~\cite{rostron2016raman}. 

\begin{figure}[h!]
    \centering
    \includegraphics[width=\textwidth]{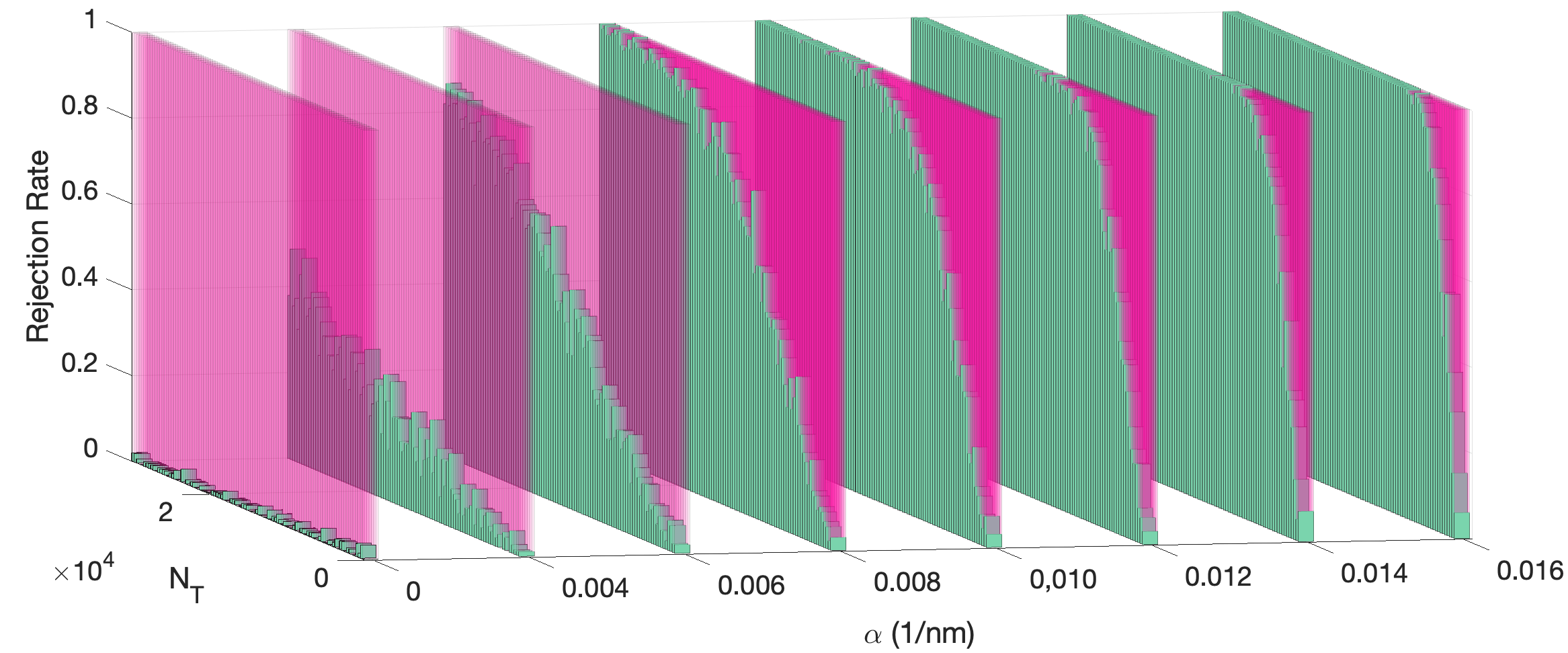}
    \caption{KS Rejection rate for different values of $\alpha$ and resources $N_T$. Green bars indicate the successful rejection. Pink bars indicate acceptance of the null hypothesis.  }
    \label{fig:sim1RR}
\end{figure}

For the first simulation, we consider as a reference a Gaussian pulse $R=\exp(-(\omega-\omega_0)^2/(2\sigma_w^2))$ centered at $\lambda_0=805$ nm, with $\sigma_{\lambda}=4$ nm, and a spectral object with transmittance $T=1-\alpha \lambda$ with $\alpha$ varying between 0 and 0.016 1/nm, as shown in panel a) o Fig. \ref{fig:sim1s}. We simulate the measured reference by considering a total of $N_R=350k$ resources and generating the measured counts by extracting random values from a Poissonian distribution centered at $N_R\cdot R$. 

The simulated signal is obtained analogously, by multiplying the signal profile by the resources interacting with the spectral object $N_T$. Since different values of $\alpha$ correspond to a different transmittivity, this will amount to a different number of detected photons depending on the spectral profile. We vary $N_T$ from 300 to 30000, and for each level of signal we randomly extract 100 simulated profiles, and perform a KST for each profile. In Fig. \ref{fig:sim1RR} we report the rejection rate normalized over the 100 simulated experiments for varying $\alpha$ and $N_T$.

\begin{figure}[b!]
    \centering
    \includegraphics[width=\textwidth]{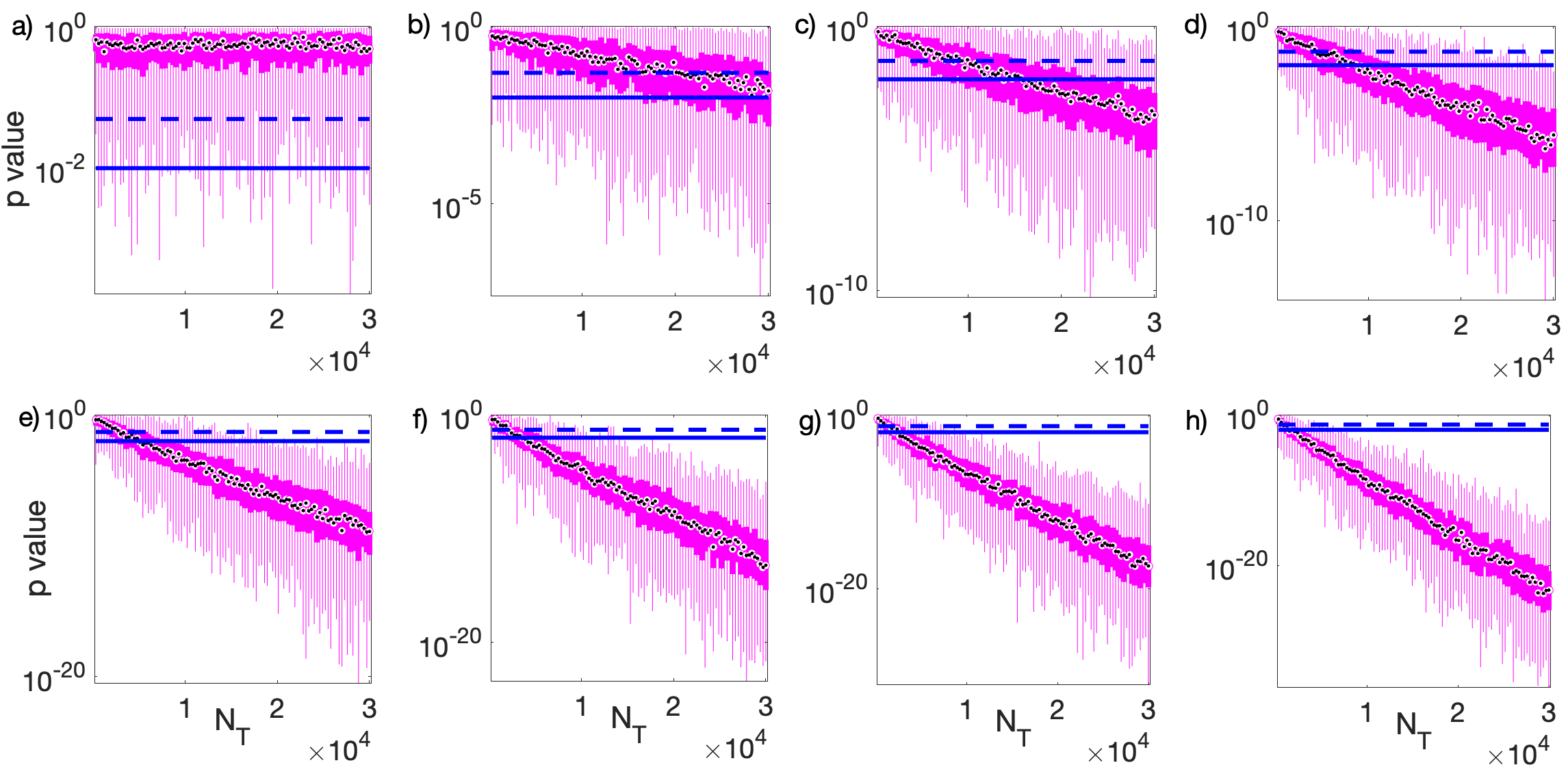}
    \caption{{\bf p values} a) $\alpha=0$ b) $\alpha=0.004$ c) $\alpha=0.006$ d) $\alpha=0.008$ e) $\alpha=0.010$ f) $\alpha=0.012$ g) $\alpha=0.014$ and h) $\alpha=0.016$. Dashed blue line: 0.05 confidence level; solid blue line: 0.01 confidence level. The box center (black dot) indicates the average value, while the box edges  (dark solid pink)  indicate the 25th and 75th percentile; the whiskers extend to all measured values.}
    \label{fig:sim1p}
\end{figure}

 When $\alpha=0$ the signal is equivalent to the reference: indeed, the rejection rate is below $3\%$ at all signal levels. This shows that even with very low signal levels, the absence of the spectral object is almost always correctly detected. The higher $\alpha$, the more different the signal distribution from the reference, and the fewer resources are needed for discriminating the two profiles. 
In Fig. \ref{fig:sim1p} we report the p-values for each $\alpha$ as a function of the number of resources $N_T$. As expected, for $\alpha=0$ the p values are well above the confidence level and as $\alpha$ increases the p value becomes smaller. 

\begin{figure}[t!]
    \centering
    \includegraphics[width=\textwidth]{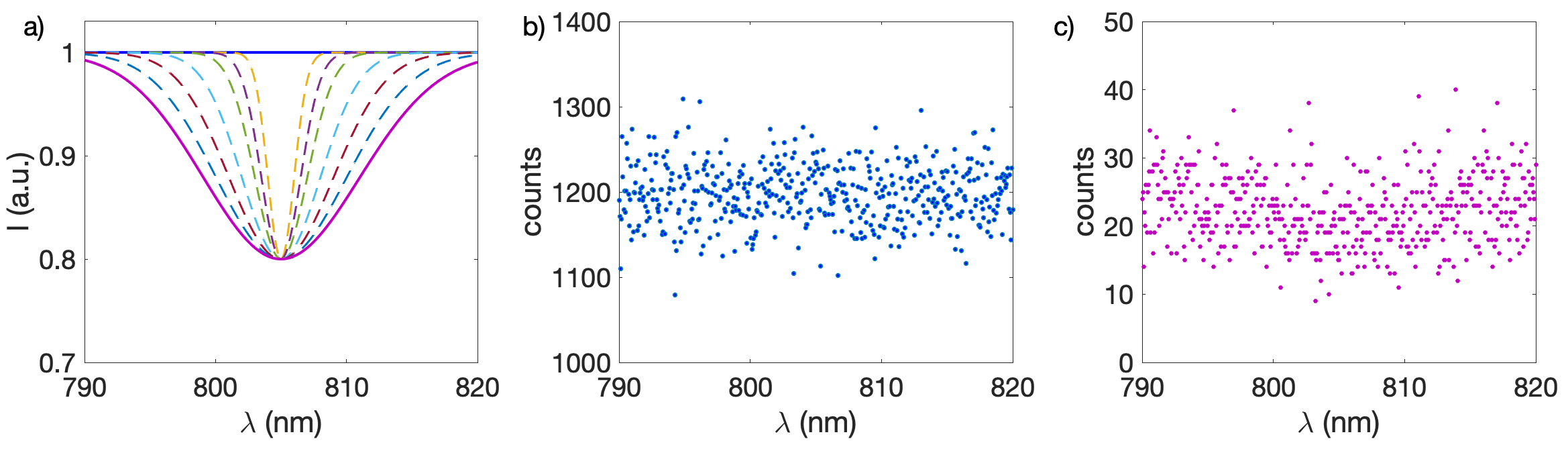}
    \caption{Simulation for broad reference and narrow absorption. a) blu: reference, dashed lines: spectral object transmission, purple: signal obtained with the yellow transmission profile.  b) simulated reference c) simulated signal for $\sigma=6$ nm for $N_T=15k$ resulting in a $100\%$ rejection rate.}
    \label{fig:sim2s}
\end{figure}

\begin{figure}[b!]
    \centering
    \includegraphics[width=\textwidth]{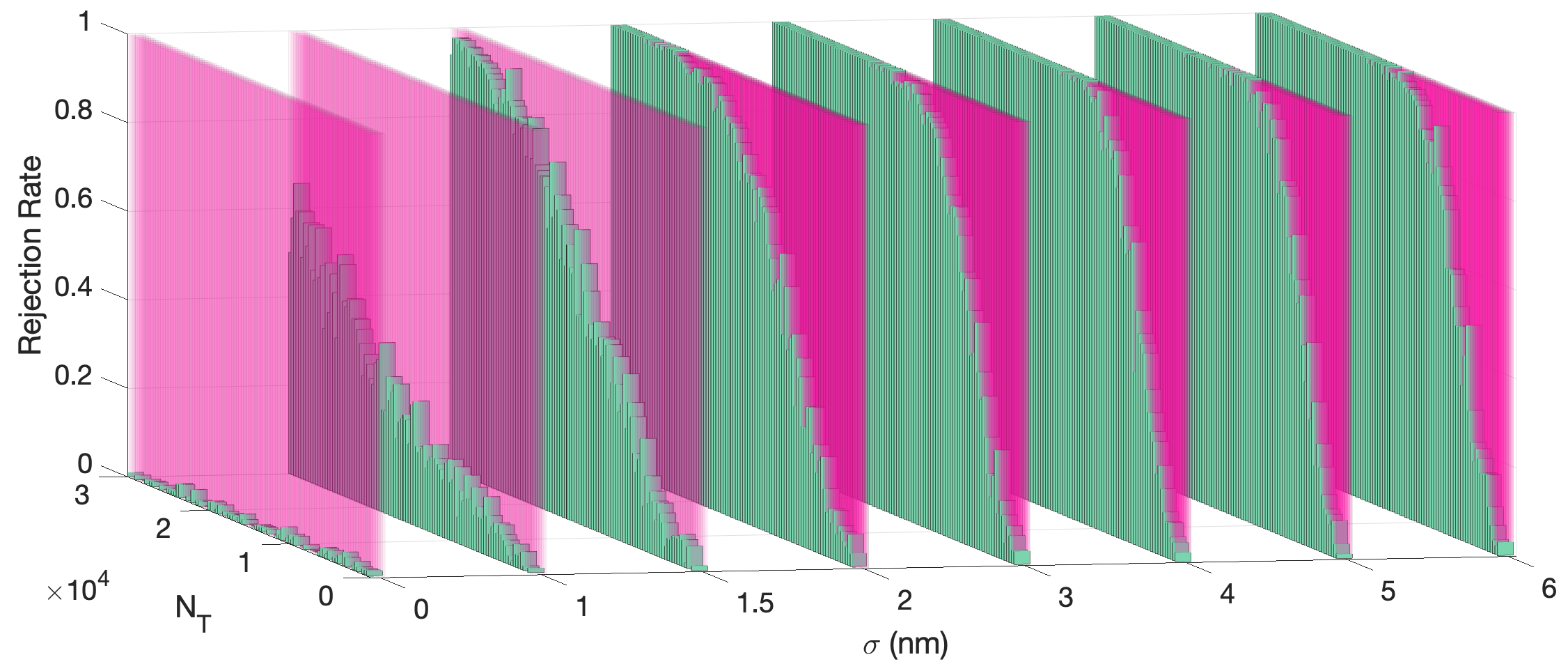}
    \caption{KS Rejection rate for different values of $\sigma$ and resources $N_T$. Green bars indicate the successful rejection. Pink bars indicate acceptance of the null hypothesis.  }
    \label{fig:sim2rr}
\end{figure}

\begin{figure}[b!]
    \centering
    \includegraphics[width=\textwidth]{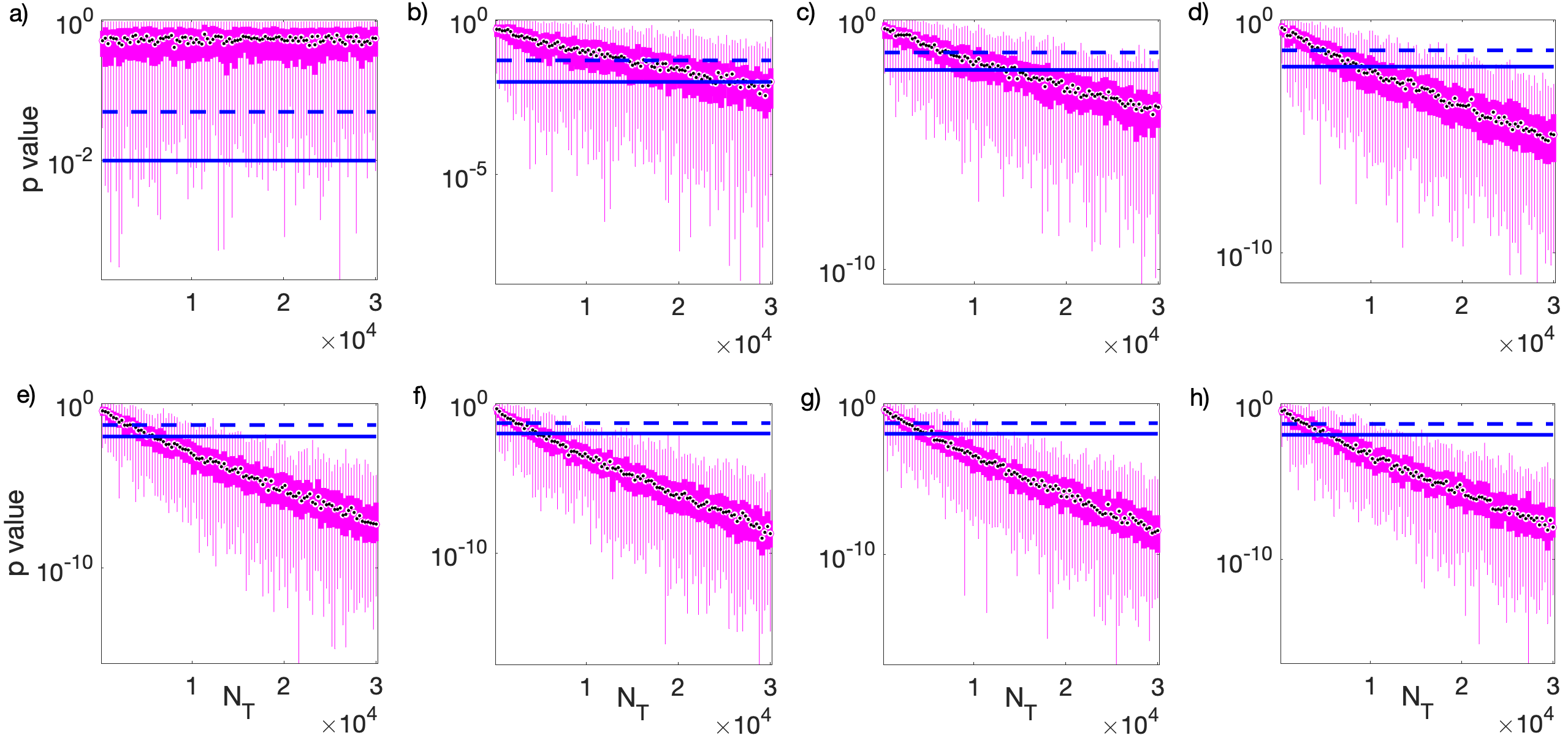}
    \caption{{\bf p values} a) $\sigma=0$ nm b) $\sigma=1$ nm c) $\sigma=1.5$ nm d) $\sigma=2$ nm e) $\sigma=3$ nm f) $\sigma=4$ nm g) $\sigma=5$ nm and h) $\sigma=6$ nm. Dashed blue line: 0.05 confidence level; solid blue line: 0.01 confidence level.The box center (black dot) indicates the average value, while the box edges (dark solid pink)  indicate the 25th and 75th percentile; the whiskers extend to all measured values.}
    \label{fig:sim2p}
\end{figure}

We now turn to the scenario in which the reference is broader than the absorption feature. 
We consider a flat reference profile and we model the transmission as $T=1- \alpha \cdot \exp(-(\lambda-\lambda_0)^2/(2\sigma^2))$. We keep $\alpha=0.2$ fixed and vary the width $\sigma$ from 0 to 6 nm. The resulting profiles are shown in panel a) of  Fig. \ref{fig:sim2s}. As before, we simulate the measured reference and the signal, however given the different profile shape, to attain the same average counts per bin for the reference we now employ $N_R=600k$ resources. The simulated reference profile is shown in panel b) of Fig. \ref{fig:sim2s}. We then simulate the transmitted profile following the same procedure described above, generating 100 simulated profiles for each level of signal and $\sigma$.In panel c) we show an example of simulated signal relative to the transmission profile with the broadest dip, for $N_T=15k$  resources. For each generated signal we perform a KST and report the obtained rejection rate in Fig. \ref{fig:sim2rr}. Even under these unfavourable conditions, dictated by a low absorption ($\alpha=0.2$) and by a narrow peak, the method achieves a satisfactory results, albeit requiring more resources than in the previous instance. In Fig. \ref{fig:sim2p} we report the obtained p-values for each $\sigma$ as a function of $N_T$.

\section{Conclusion}

In this article we have explored the use of a quantum ghost spectrometer for discriminating the presence of an absorbing spectral object. Using a Kolmogorov-Smirnov test we have been able to asses the presence of an object by exploiting a limited amount of resources in an efficient way. We have demonstrated our technique with experiments on two distinct samples, as well as with simulations extending the experimental results to typical spectral regimes. 

Our technique provides a viable route for fast discrimination of a spectral object in all the examined condition. In particular, the more different the spectral profile after the absorption, the lesser resources are needed for a successful discrimination. This makes our technique an optimal solution when dealing with systems that strongly affect the transmission resulting in exceedingly low signal rates, which, coincidentally, are those that would otherwise require more effort for reconstructing the full lineshape. Moreover, our approach offers great spectral tunability as it can take advantage of the non-degenerate measurements, enabling the investigation in otherwise hardly accessible spectral regimes. 

Our results can be extended in different directions. 
Solutions taken from  Fuzzy Logic \cite{chemosensors10080295} or Machine learning algorithms \cite{Moodley23} can benefit the hypothesis testing approach. In this respect, Machine Learning has been employed for classification of non-spectral features \cite{Giordani20}. This suggests that the efficiency of the method can be further optimised. As for the extension of the capabilities, incorporating other degrees of freedom, notably the spatial domain or the polarisation, would be particularly helpful in determining not only the presence of a threat but also its position and size.

\section*{Acknowledgements}
This work was supported by the NATO Science for Peace and Security (SPS) Programme, project HADES (id G5839). We acknowledge helpful discussions with Simone Santoro.


\bibliography{report}   

\begin{thebibliography}{10}

\bibitem{pavia2014introduction}
D.~L. Pavia, G.~M. Lampman, G.~S. Kriz, {\em et~al.}, {\em Introduction to
  spectroscopy}, Cengage learning  (2014).

\bibitem{koral2021thz}
C.~Koral, G.~Papari, and A.~Andreone, ``Thz spectroscopy of advanced
  materials,'' {\em Terahertz (THz), Mid Infrared (MIR) and Near Infrared (NIR)
  Technologies for Protection of Critical Infrastructures Against Explosives
  and CBRN} , 253--273  (2021).

\bibitem{hammes2005spectroscopy}
G.~G. Hammes, {\em Spectroscopy for the biological sciences}, John Wiley \&
  Sons  (2005).

\bibitem{adriaens2021spectroscopy}
M.~Adriaens and M.~Dowsett, {\em Spectroscopy, diffraction and tomography in
  art and heritage science}, Elsevier  (2021).

\bibitem{appenzeller2012introduction}
I.~Appenzeller, {\em Introduction to astronomical spectroscopy}, vol.~9,
  Cambridge University Press  (2012).

\bibitem{schaefer1998spectroscopic}
K.~Schaefer, {\em Spectroscopic atmospheric environmental monitoring
  techniques}, Society of Photo-Optical Instrumentation Engineers, Bellingham,
  WA  (1998).

\bibitem{wagatsuma2021spectroscopy}
K.~Wagatsuma, {\em Spectroscopy for Materials Analysis: An Introduction},
  Springer  (2021).

\bibitem{Mukamel20}
S.~Mukamel, M.~Freyberger, W.~Schleich, {\em et~al.}, ``Roadmap on quantum
  light spectroscopy,'' {\em Journal of Physics B: Atomic, Molecular and
  Optical Physics} {\bf 53}, 072002  (2020).

\bibitem{Kalachev_2008}
A.~A. Kalachev, D.~A. Kalashnikov, A.~A. Kalinkin, {\em et~al.}, ``Biphoton
  spectroscopy in a strongly nondegenerate regime of spdc,'' {\em Laser Physics
  Letters} {\bf 5}, 600  (2008).

\bibitem{Chan09}
K.~W.~C. Chan, M.~N. O'Sullivan, and R.~W. Boyd, ``Two-color ghost imaging,''
  {\em Phys. Rev. A} {\bf 79}, 033808  (2009).

\bibitem{Kalashnikov16}
D.~Kalashnikov, A.~Paterova, S.~Kulik, {\em et~al.}, ``Infrared spectroscopy
  with visible light,'' {\em Nature Photonics} {\bf 10}, 98--101  (2016).

\bibitem{Paterova20}
P.~Anna, V., M.~Sivakumar, M., Y.~Hongzhi, {\em et~al.}, ``Hyperspectral
  infrared microscopy with visible light,'' {\em Science Advances} {\bf 6}(44),
  eabd0460  (2020).

\bibitem{PhysRevX.4.011049}
D.~A. Kalashnikov, Z.~Pan, A.~I. Kuznetsov, {\em et~al.}, ``Quantum
  spectroscopy of plasmonic nanostructures,'' {\em Phys. Rev. X} {\bf 4},
  011049  (2014).

\bibitem{Aspden15}
R.~S. Aspden, N.~R. Gemmell, P.~A. Morris, {\em et~al.}, ``Photon-sparse
  microscopy: visible light imaging using infrared illumination,'' {\em Optica}
  {\bf 2}, 1049--1052  (2015).

\bibitem{Pepe16}
F.~V. Pepe, F.~Di~Lena, A.~Garuccio, {\em et~al.}, ``Correlation plenoptic
  imaging with entangled photons,'' {\em Technologies} {\bf 4}(2)  (2016).

\bibitem{Ryczkowski16}
P.~Ryczkowski, M.~Barbier, A.~T. Friberg, {\em et~al.}, ``Ghost imaging in the
  time domain,'' {\em Nature Photonics} {\bf 10}, 167--170  (2016).

\bibitem{lemos14nat}
G.~B. Lemos, V.~Borish, G.~D. Cole, {\em et~al.}, ``{Quantum imaging with
  undetected photons},'' {\em Nature} {\bf 512}(7515), 409--412  (2014).

\bibitem{Moreau18}
P.-A. Moreau, E.~Toninelli, T.~Gregory, {\em et~al.}, ``Ghost imaging using
  optical correlations,'' {\em Laser \& Photonics Reviews} {\bf 12}(1), 1700143
   (2018).

\bibitem{chierici20}
A.~Chierici, ``2nd scientific international conference on cbrne sicc series:
  2020: book of abstracts,'' in {\em 2nd scientific international conference on
  CBRNe SICC series},  1--172, TAB  (2020).

\bibitem{sullivan10pra}
M.~N. O'Sullivan, K.~W.~C. Chan, and R.~W. Boyd, ``Comparison of the
  signal-to-noise characteristics of quantum versus thermal ghost imaging,''
  {\em Phys. Rev. A} {\bf 82}, 053803  (2010).

\bibitem{bennink04prl}
R.~S. Bennink, S.~J. Bentley, R.~W. Boyd, {\em et~al.}, ``Quantum and classical
  coincidence imaging,'' {\em Phys. Rev. Lett.} {\bf 92}, 033601  (2004).

\bibitem{padgett17ptrsa}
M.~J. Padgett and R.~W. Boyd, ``An introduction to ghost imaging: quantum and
  classical,'' {\em Philosophical Transactions of the Royal Society A:
  Mathematical, Physical and Engineering Sciences} {\bf 375}(2099), 20160233
  (2017).

\bibitem{PhysRevA.105.013506}
A.~Chiuri, I.~Gianani, V.~Cimini, {\em et~al.}, ``Ghost imaging as loss
  estimation: Quantum versus classical schemes,'' {\em Phys. Rev. A} {\bf 105},
  013506  (2022).

\bibitem{ghorbani2019mahalanobis}
H.~Ghorbani, ``Mahalanobis distance and its application for detecting
  multivariate outliers,'' {\em Facta Universitatis, Series: Mathematics and
  Informatics} , 583--595  (2019).

\bibitem{Silva92}
D.~S. Sivia and C.~J. Carlile, ``Molecular spectroscopy and bayesian spectral
  analysis—how many lines are there?,'' {\em J. Chem. Phys.} {\bf 96}
  (1992).

\bibitem{Paris19}
I.~Maffeis, S.~Koudia, A.~Gharbi, {\em et~al.}, ``Process estimation in qubit
  systems: a quantum decision theory approach,'' {\em Quantum Information
  Processing} {\bf 18}  (2019).

\bibitem{scott}
C.~Scott and R.~Nowak, ``A neyman-pearson approach to statistical learning,''
  {\em IEEE Transactions on Information Theory} {\bf 51}(11), 3806--3819
  (2005).

\bibitem{razali2011power}
N.~M. Razali, Y.~B. Wah, {\em et~al.}, ``Power comparisons of shapiro-wilk,
  kolmogorov-smirnov, lilliefors and anderson-darling tests,'' {\em Journal of
  statistical modeling and analytics} {\bf 2}(1), 21--33  (2011).

\bibitem{Perkampus12}
H.-H. Perkampus, {\em UV-VIS Spectroscopy and Its Applications}, Springer
  Berlin, Heidelberg  (2012).

\bibitem{Griffiths07}
P.~R. Griffiths, J.~A. De~Haseth, and W.~J. D., {\em Fourier Transform Infrared
  Spectrometry}, Wiley  (2007).

\bibitem{rostron2016raman}
P.~Rostron, S.~Gaber, and D.~Gaber, ``Raman spectroscopy, review,'' {\em laser}
  {\bf 21}, 24  (2016).

\bibitem{chemosensors10080295}
F.~Angelini, S.~Santoro, and F.~Colao, ``Chemical identification from raman
  peak classification using fuzzy logic and monte carlo simulation,'' {\em
  Chemosensors} {\bf 10}(8)  (2022).

\bibitem{Moodley23}
C.~Moodley, A.~Ruget, J.~Leach, {\em et~al.}, ``Time-efficient object
  recognition in quantum ghost imaging,'' {\em Advanced Quantum Technologies}
  {\bf 6}(2), 2200109  (2023).

\bibitem{Giordani20}
T.~Giordani, A.~Suprano, E.~Polino, {\em et~al.}, ``Machine learning-based
  classification of vector vortex beams,'' {\em Phys. Rev. Lett.} {\bf 124},
  160401  (2020).

\end{thebibliography}
\bibliographystyle{spiejour}   



\end{spacing}
\end{document}